# Service Function Chaining in 5G & Beyond Networks: Challenges and Open Research Issues

Hajar Hantouti, Nabil Benamar, and Tarik Taleb

## Abstract

Service Function Chaining (SFC) is a trending paradigm, which has helped to introduce unseen flexibility in telecom networks. Network service providers, as well as big network infrastructure providers, are competing to offer personalized services for their customers. Hence, added value services require the invocation of various elementary functions called Service Functions (SFs). The SFC concept composes and imposes the order in which SFs are invoked for a particular service. Emerging technologies such as Software Defined Networking and Network Function Virtualization support the dynamic creation and management of SFC. Even though SFC is an active technical area where several aspects were already standardized and many SFC architecture flavors are currently deployed, yet some challenges and open issues are still to be solved. In this paper, we present different research problems related to SFC and investigate several key challenges that should be addressed to realize more reliable SFC operations.

## Introduction

Along with the incremental demands on networking services, customers require increasingly advanced, customized, and sometimes sophisticated services. Complex services require composing a set of elementary Service Functions (SFs) to satisfy the technical clauses depicted in service level agreements. A video streaming service, as such, can vary depending on the traffic location, customer preferences, network state, and other policies. Yet, it is a tedious task to stitch SFs together to compose added value services.

Service Function Chaining (SFC) is a networking architecture that creates a service chain of connected network services. Traffic can be bound to Service Chains (SC) by identifying traffic types using, for example, Virtual Local Area Networks (VLANs) or tunnels. However, such methods are complex and require error-prone configurations due to their topological adherence constraints.

Recently, important industrial and research efforts have been undertaken for the sake of dynamic SFC schemes [1]. Moreover, standardization bodies such as ETSI and IETF published documents such as the RFC 7665 [2] that specifies a reference SFC data plane architecture making use of separate transport and service encapsulation, while RFC 8300 [3] defines Network Service Header (NSH) as the SFC service encapsulation.

To meet the requirements of a flexible and programmable SFC, networking technologies such as Software Defined Networking (SDN) and Network Function Virtualization (NFV) can be designed for this purpose. By separating the control plane from the data plane, SDN presents a centralized programming and control tool using dedicated interfaces and communication protocols to the underlying networking devices in the data plane. Furthermore, NFV promotes the development of Virtual Network Functions (VNFs or virtual SFs) to be deployed on commodity hardware.

Though SFC aims to deploy flexible and complex services, we currently lack a clear understanding of the solutions for several SFC problems. Very few articles focus on research challenges and open issues in the SFC research area. For example, John *et al*. [4] presented research directions for SFC, motivating the research for dynamic SFC. John *et al*. discussed research challenges related to SC description, programming, deployment, and debugging. Later, Medhat *et al*. [5] presented a state of the art of SFC proposals and highlighted the related challenges, such as Traffic Steering (TS), QoS, placement and resource allocation. Recently, Zhang *et al*. [6] discussed the SFC architecture, challenges and opportunities for enabling efficient SFC by integrating SDN and NFV. Zhang *et al*. focused on three research areas: service modeling, resource allocation and TS. However, more issues remain unsolved for more efficient and reliable SFC operations. The current paper identifies several pending research problems in SFC through different stages of service deployment. Our goal is to present a comprehensive list of challenges that covers a wide range of SFC research areas: SFC management and orchestration, SFC composition, path selection, placement of SFs, service allocation and provisioning, TS, QoS and security. Moreover, we present some open research issues resulting from the latest developments in SFC that have not been discussed in previous works, as well as remaining SFC challenges.

The remainder of this paper is organized as follows. First, we introduce the concept of SFC with a detailed presentation of its architectural components and we briefly describe the technologies supporting SFC. Afterward, we highlight the different research problems related to SFC along

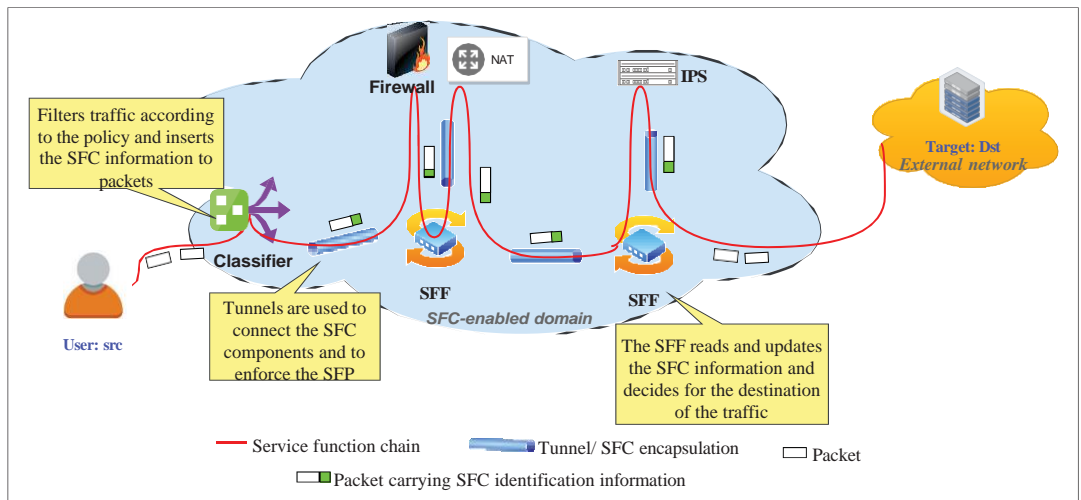

FIGURE 1. Service Function Chaining use case.

with the remaining relevant challenges and open issues. Finally, we conclude the paper.

## SERVICE FUNCTION CHAINING: SCOPE
### SERVICE FUNCTION CHAINING: THE CONCEPT AND BENEFITS

SFC has become part of mobile networks, data centers, and broadband networks. When SFC is deployed in the SDN/NFV context, it allows for composing customized services and supports fine granular policies. SFC permits to avoid strong adherence to the underlying physical topology and provides better deployment flexibility. Moreover, it ensures a dynamic service inventory, whereby SFs can be added or removed without breaking the chain. Currently, SCs are seen as graphs of SFs instead of linear SCs.

SFC is defined in RFC7665[2], and refers to the definition, instantiation and steering of the network traffic through an ordered list of SFs. As a basic example, a chain may be composed of {Firewall, NAT and IPS} (Fig.1).

In Fig.1, the traffic issued from "user:src" and destined to "target:dst" is subject to a classification process at the ingress of the network, and then enters the chain. First, the traffic is directed to the Firewall, then forwarded to the NAT and passed to the IPS afterward. In fact, such a use case is common in networks to enforce security. The added value of SFC here is the rationalization of the way the SFs are connected and used.

SFC use cases represent specific network policies, customer strategies and users' preferences. As such, SFC may offer different QoS for different customer profiles (e.g., premium or basic) or customer media (mobile phone, laptop) and thus optimize the networking parameters (e.g., bandwidth, latency) accordingly. The SFC concept may also be used to optimize the video streaming cost and network resources, as well as for some personalized parental control and security policies (e.g., steering suspicious/voluminous traffic to a scrubbing center).

### SERVICE FUNCTION CHAINING: THE ARCHITECTURE

The main architectural concept in SFC is the separation of the logical SFC overlay and the data plane. As stated in RFC 7665 [2] that describes the specification of an SFC architecture, the packet handling operations are separated from the realization of Service Function Paths (SFP). In other words, SFPs are realized in an abstraction of the packet handling operations (e.g., Packet forwarding). Moreover, the SFC architecture is independent from the underlying network topology, which means that topological changes do not affect SFC operations.

The SFC architecture defines some components that are responsible for a set of traffic operations and they are placed along the SFP. It includes classifiers, service function forwarders, service function nodes and proxies when needed.

**Classifier:** The classifier (CL) is responsible for the classification of the traffic. The process of classification permits filtering different traffic types according to policy profiles. To avoid re-classification at every SFC element, an identification process takes place for the next SFC element to process packets based on the result of the ingress classification process. Usually, Service Path Identifiers (SPIs) are inserted in packets [2].

**Service Function Forwarder (SFF):** It is responsible for forwarding traffic between SFC components over the SFC overlay. The SFC forwarding operation is based on flow identifiers of traffic types or by matching rules in SFFs. The forwarding operations result in selecting the next SFC element in the SFP. In case a path identifier and/or metadata are added to the packets, an encapsulation process takes place to ensure connection and delivery between SFC elements [2].

**SFC Proxy:** The proxy is placed between the SFF and SFs that are not SFC-aware. Its role is to enable SFC communication between the SFs and SFF, in other words, to integrate the SF in the service chain. It implements SFC functions, such as adding or consuming metadata on behalf of the SFs.

**Service Function (SF) Node:** The SF node is where the service function is deployed. It can be of different forms, hardware, or virtualized node. Also, it can host one or more SFs.

### TECHNOLOGIES PROMOTING SFC

Different standardization bodies are promoting the SFC deployment, namely IETF, ETSI and ONF (among others). The IETF has mainly contributed in defining the SFC problem, architecture, Oper-

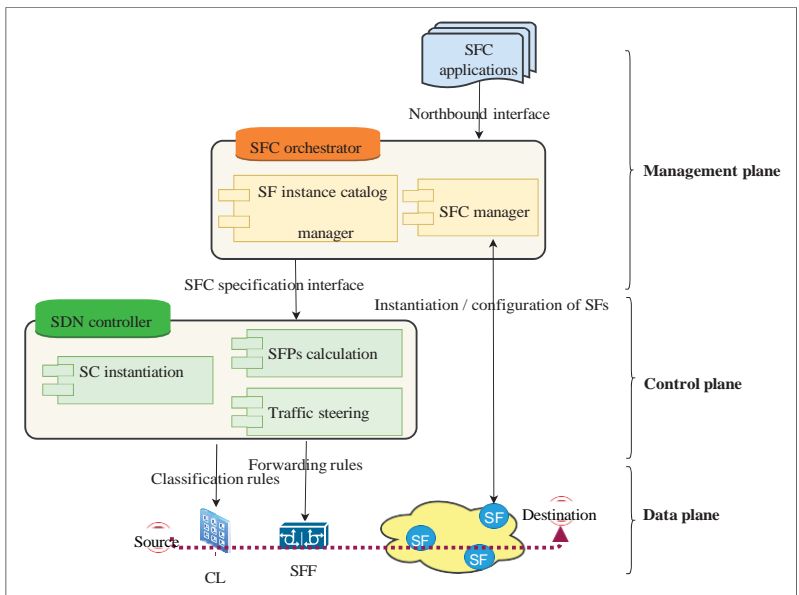

FIGURE 2. Service function chaining in an SDN enabled network.

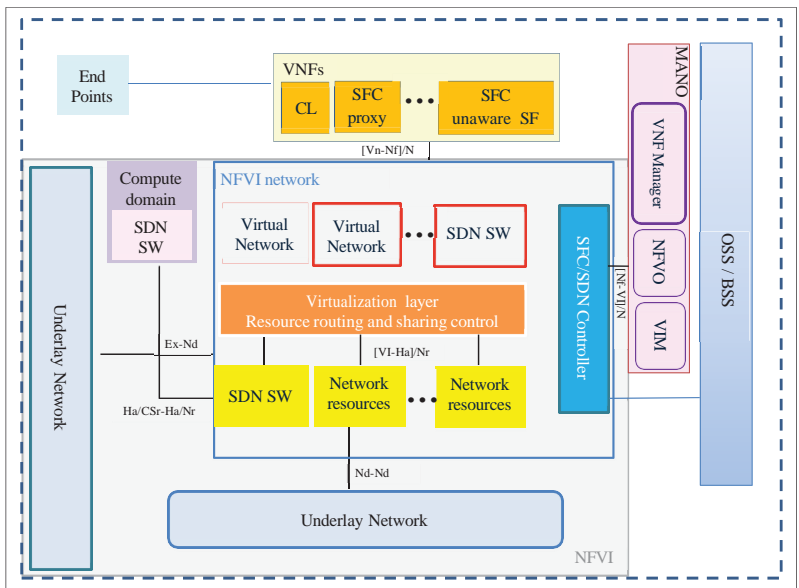

FIGURE 3. Service function chaining in an SDN & NFV enabled network.

Fig.2 describes a SFC deployment in the context of SDN. SDN permits to dynamically manage network operations in SFC. It allows programming the classifiers and the forwarders to enable reactive or proactive installation of flow rules. Thus, it adds flexible and dynamic traffic forwarding for SFC.

**Network Function Virtualization:** ETSI has recently standardized the NFV architecture. NFV claims to improve flexibility and efficiency in deploying VNFs. The SDN controllers and SFC components can all be virtualized and orchestrated by the NFV-orchestrator. NFV is assumed to accomplish scalability and underlying topology abstraction. Thus, leveraging NFV for SFC can improve the flexibility and reduce costs for communication and investment for physical SFs.

There are different deployments of NFV and SDN for SFC. The SDN controller can be part of the NFV Infrastructure (NFVI), and maybe part of the Virtual Infrastructure Manager (VIM) [7]. Fig. 3 describes an example of SFC in an SDN&NVF environment whereby the SDN controller is part of the NFVI and connects with the VIM. The SFC elements (i.e., SFs, CL, proxy, SFF) can be virtualized as well as the SDN controller. The SFFs are represented by the SDN switches (SDN SW), and can be deployed either as VNF instances, network resources in the NFVI, or as hardware appliances, whereas the NFV Orchestrator (NFVO) interfaces with the VNF managers, deploys the VNFs and manages the network resources.

## Key Challenges and Open Issues

This section identifies several pending SFC research challenges, at different design and implementation levels. Specifically, the research challenges pertain to SFC enabled in an SDN and/or NFV environment. Table 1 presents a summary of the current SFC challenges, classified into different SFC research areas.

### Service Chains Composition, SFC Path Selection and Placement of Service Functions

The SFC composition problem is directly related to the SF placement and path selection problems. These three problems contribute to defining the service chains and selecting the optimal SFC path and SFs. The service chain deployment process can be achieved in different interchangeable steps. Different algorithms can calculate the chain graph, find the best path, and accordingly place SFs.

First, the SFC composition problem refers to the operations involved in translating SFCs from an abstract layer, defined during the design time, to a concrete set of SFs (Fig.4.(a)). The goal of SFC composition remains in the abstraction of SFCs, in order to implement complex services without worrying about implementation details. As a result of the composition process, the accurate SFs, corresponding to a service chain, can be instantiated.

Composing SFCs and mapping them to physical resources is still challenging, and this is despite the different composition and mapping algorithms proposed in the literature. Several criteria make the composition and mapping to SFs a challenging problem. For instance, composing SFCs based on service constraints, actual networking state, infrastructure capabilities and subscribers' prefer-ation Administration and Maintenance (OAM), traffic steering techniques and new protocols. While ONF proposed an SFC deployment using SDN and Openflow, based on the IETF work, ETSI focused on deploying SFC in NFV, and the integration of SDN controllers.

**Software Defined Networking:** SDN enables network programmability, assuming abstraction techniques used by the SDN computation logic to dynamically enforce policies as a function of the nature of the service to be delivered.

As described in [2], SFC relies upon a control plane and policy constructs. There is an upper layer above the control plane composed of business applications, referred to as the management layer. This layer sends applications requests to the control plane that translates the policy requirements into forwarding rules. The deployment of SFC in an SDN-enabled environment allows creating, managing and controlling chains by software, in an abstraction of the underlying network topology.

ences is not a straightforward problem. In order to satisfy some constraints, other constraints may be violated; e.g., re-using SFs can guarantee cost reduction but may result in longer paths.

Furthermore, in an SFC environment, different types of SFs can be deployed (i.e., hardware appliance, VM image, packet I/O driver, container process). This heterogeneity introduces interoperability constraints. Deploying virtual SFs of the same type can reduce deployment cost since different SFs can be combined in the same node. In the heterogeneous scenarios, the use of different types of SFs can increase the cost. Therefore, the composition of chains in heterogeneous environments is also challenging.

Second, service chains can be deployed using unique paths or multiple paths. It depends on the policy profiles set in the management plane. The path selection problem for SFC results in selecting different paths to the same chain according to the QoS and infrastructure/operator policy (Fig.4.(c)). Indeed, differentiating paths allows for satisfying different SFC constraints. For example, some policy profiles can prioritize cost minimization along with increasing latency and link bandwidth. Other cases can prioritize delivery time or shortest path without considering VMs (nodes that embed VNFs, usually nodes are Virtual Machines) usage. However, a multi-criteria path selection is a challenging problem and the current methods for path selection based on shortest path selection do not guarantee QoS.

Another issue consists in the fact that once paths are calculated for given chains, it is challenging to update the SFP in real-time and during the service delivery based on current network state and environment changes.

Third, the placement problem aims to determine the optimal SF locations, in other words, mapping SFs to nodes. The objective is to provide high network performance and efficient resource utilization (Fig.4.(b)). Although a lot of work on SFs placement has been carried out, further research is needed for customizing placement constraints taking into consideration the delivery time, the generated cost, the preferences of subscribers, and the properties of the infrastructure. Moreover, placement algorithms should consider realistic cases where some SFs cannot coexist in the same node because of conflict or restriction. Though combining some SFs in a node can be optimal (e.g., bandwidth consumption) for specific chains, it may not be optimal for other SFCs. Considering these metrics makes the placement problem even more challenging.

### SERVICE ALLOCATION AND PROVISIONING

Resource allocation remains one of the main challenges of NFV and SFC. It directly relates to the placement and composition problems (Fig.4.(b)). Although the dynamic resource provisioning allows allocating resources for VNFs when required, the challenge of resource sharing between VNFs is present. Therefore, there is a competition among VNFs for global resources. Thus, some VNFs may run out of resources, causing VNF failures, and ultimately impacting the overall service delivery of the chain. To ensure an acceptable level of network performance, resources may be wasted, resulting in resources idleness issue. In fact, while some SF instances become a bottleneck in the process of SFC due to reduced resources, other SF instances may remain idle, occupying unnecessary resources. Thereby, the resource utilization ratio should be considered in the SFC orchestration and resource allocation.

| SFC research areas | Current challenges |
|---|---|
| Service chain composition | <ul><li>Mapping SCs to physical resources</li><li>Some criteria makes the composition challenging: services constraints, actual networt state, infrastructure capabilities, subscribers preferences</li><li>Some criteria can be contradictory</li><li>Heterogeneity of SFs types and environments introduces interoperability constraints</li></ul> |
| Path selection | <ul><li>Multi-criterea path selection: latency, link bandwidth, delivery time, VMs usage, delivery time …</li><li>Updating the SFP in real-time service delivery based on current network state and environment changes</li></ul> |
| Placement of SFs | <ul><li>Taking into consideration: delivery time, generated cost, infrastructure properties and subscribers preferences</li><li>In some cases, SFs cannot coexist in the same node because of conflict or restriction</li></ul> |
| Service allocation and provisioning | <ul><li>Resource sharing between VNFs, competition for global resources</li><li>Resource idleness</li></ul> |
| Orchestration | <ul><li>Adaptation of SFPs and VNF instances to network requests, network state.</li><li>Traffic distribution variation in time in the network</li><li>Dynamic discovery of SFs</li><li>Adaptation to real-time changes to ensure reliable SFC delivery</li><li>Reducing human interference</li><li>Automating management operations</li></ul> |
| Traffic steering | <ul><li>Support of multiple TS protocols/techniques for various scenarios</li><li>Size of forwarding state and reources constraints</li><li>Consecutive modifications on packets headers results in additional delay and can raise security concerns</li><li>Encapsulation protocols can introduce MTU issues and inconsistency</li><li>Support for hybrid symmetry</li><li>Interoperability between TS techniques in heterogeneous networks</li><li>Techniques to allow translation betwwen different TS methods is needed</li><li>TS for SFC in multi-tenant networks</li><li>Support of SFC headers and encapsulation protocols by the SFs</li></ul> |
| QoS and QoE | <ul><li>QoS ane QoE assessement and visualization</li><li>The combination of networks metrics, sometimes contradictory metrics should be combined</li><li>Respecting the QoE while maximizing the QoS</li><li>Dynamically analyzing the QoS and QoE.</li><li>Adapting the QoS and QoE to network changes, and to highly dynamic environments</li></ul> |
| Security | <ul><li>TS and encapsulation protocols must be filtered at the boundaries of the SFC domain along with continuous audits</li><li>Authentification and checkups must be applied before the classification process</li><li>Tests for trusted devices ( CLs, SFFs, SFs ...)</li><li>Some infrastructure policies block SFC operations</li></ul> |

TABLE 1. Summary of SFC research areas and current challenges.

### TRAFFIC STEERING

Problems in TS are mainly related to traffic forwarding operations (Fig.4.(d)). Indeed, SDN enhances forwarding flexibility and control. However, relevant challenges are still remaining for TS in SFC. Mainly, the size of forwarding state (i.e. the set of flow entries in all the flow tables) depends on the number and the type of flow rules saved in the forwarding devices. This can be a limiting factor, particularly for devices with limited memory capacity. Due to various reasons, the forwarding state may increase (e.g., infrastructure size, number of requests, number of chains and SFs), which leads to scalability issues. To overcome this problem, some SFC solutions store the forwarding state in the controller. The original OpenFlow specifications called for this behavior, but it does not work in practice. As a result, the amount of control traffic exchanged between the SDN controller and devices increases, incurring communication overhead and wasting processing resources.

In order to enforce SFP, a TS method may insert, remove or modify packet headers. The consecutive operations on the headers result in some additional latency and processing load. Various encapsulation protocols can be used for SFC. Encapsulation increases the packet size and introduces MTU issues, mainly when multiple headers are added (i.e. network and SFC encapsulations are used) or the full SFP is encoded in the packet (e.g., source routing techniques). Intuitively, the packet size impacts the end-to-end delivery time and creates overhead. Furthermore, some encapsulation protocols are not supported by SFs. In this case, proxies are used between SFs and SFFs to enable SFC overlay communication, accordingly inducing additional complexity, resources allocation and some latency.

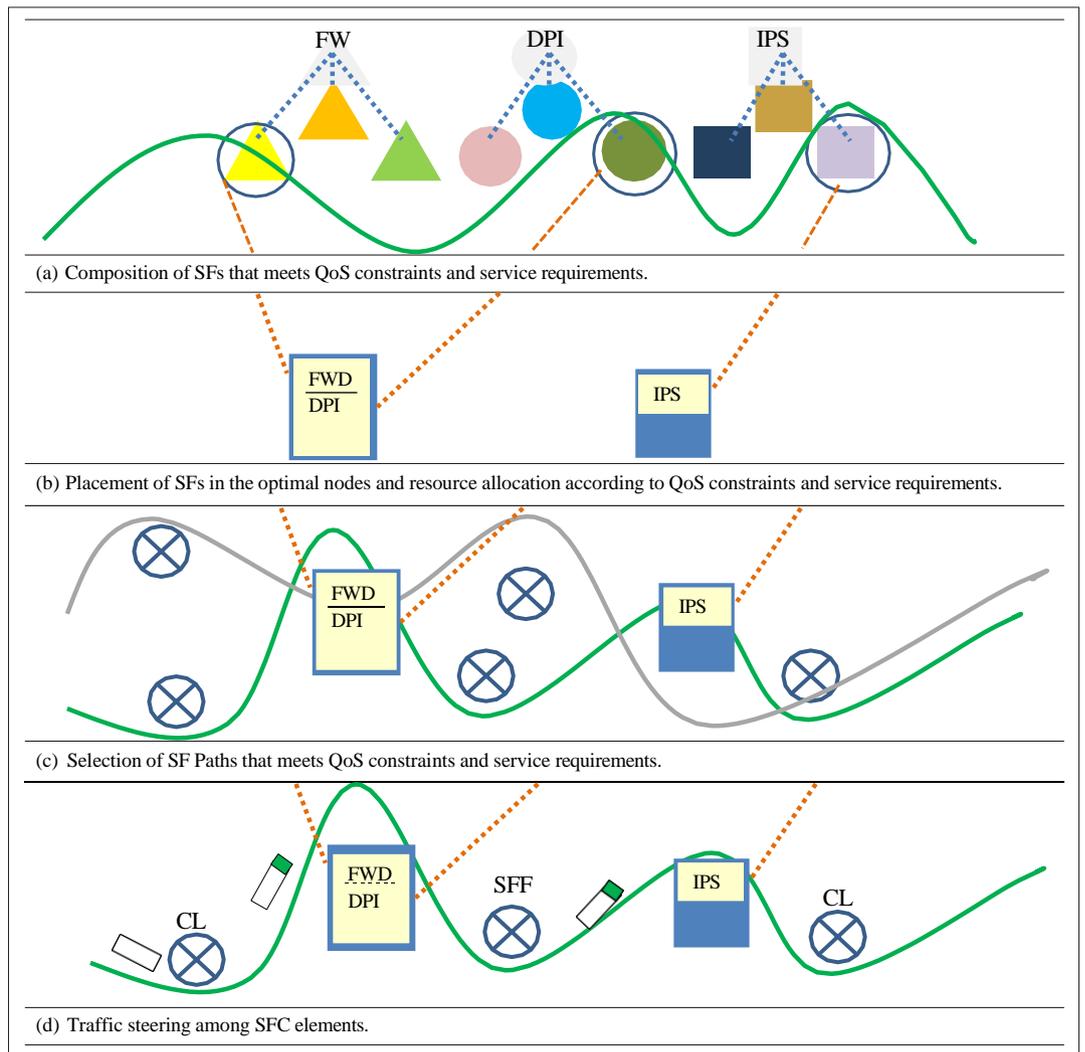

FIGURE 4. SFC workflow including different SFC stages.

SFC traffic symmetry is another challenging problem, especially in highly dynamic networks whereby the SFC underlay (the data plane or network topology and devices) changes so often. Moreover, in case of some SFs requiring symmetry, the returning traffic should pass by the same SF instance (e.g., TCP proxy, optimizer), thus the adjacent classifier must ensure that the reverse packets traverse the same path. Even though there exist some solutions to ensure traffic symmetry, they all have limitations. Also, the partial symmetry in the chains is not considered yet; the chains are defined as symmetric or asymmetric, while in fact, the overall chain symmetry depends on individual symmetry requirements of SFs. Some SFs require reverse traffic to pass through it while other SFs do not. Yet, deploying partial symmetry chains, considering SFs symmetry requirements and network changes, is challenging.

The SFC design separates the SFC layer and transport layer (routing/forwarding plane). This separation improves the interoperability as long as operators use different transports at different parts of their network, which preserves the path identification and metadata. Thus, the choice of the right TS method that does not modify the transport schemes remains the primary guarantee to interoperability. In a heterogeneous network composed of IP and MPLS parts, for example, the SFC TS method should not depend on IP nor MPLS. Indeed, a traffic steering based on IP options, IPv6 extension headers or MPLS labels is not recommended in that case. For interoperability purposes, transport-independent protocols, such as NSH, could be used. Techniques to allow translation between the different TS protocols are required. To the best of the authors' knowledge, there are no such implementations available yet.

SFC can be supported for multi-tenant networks in different mechanisms. The impact of the used mechanism consists in the amount of states in SFFs. Multi-tenancy can be supported using separate SFs and separate SFPs for different tenants, or if SFs support multi-tenancy, the tenant identifier can be carried with the SFC information (e.g., as metadata in the case of NSH protocol). The multi-tenant network may have commitments for tenants (e.g., delay, bandwidth). Nevertheless, it is challenging to offer another type of traffic differentiation for the different QoS commitments for tenants. Some techniques, such as combining the edge policing and providing resources for different DiffServ classes, can be used.

### SFC ORCHESTRATION

Several research efforts have been published recently, proposing algorithms for instantiating VNF instances according to different metrics and end-to-end service requirements. Research in this area is active to provide optimal VNF creation for SFC. However, several issues remain unsolved in SFC orchestration. This includes the adaptation of SFPs and VNF instances to network requests, network state and traffic distribution variation in time, as well as the dynamic discovery of SFs. Usually, changes in the network topology generate errors and inconsistency. Thus, there is a big challenge of adaptation to real-time changes to ensure reliable SFC delivery. Especially in large-scale networking infrastructures, the environment changes require continuous reconfigurations. Further research in automation for reducing human interference is needed to flexibly adapt to environment changes.

Although the next-generation networking technologies, namely SDN and NFV, promote network programmability and enhance service delivery, yet they increase the overall management and operation complexity. Therefore, there is a need for automating management and operations. The zero-touch automation initiative has been launched by the ETSI Zero touch network and Service Management (ZSM) Industry Specification Group (ISG) to study use cases and proof of concepts and to analyze the related challenges. At the time of writing this article, ZSM is a new issue where ZSM requirements, architecture, ZSM landscape and means of automation are the current work items. The goal of ZSM is to enable a framework that allows for agile and efficient automation management of future networks [8].

### SFC MANAGEMENT

The management of different SFC operations requires a diverse high-level application to promote flexibility. However, the focus in research is on the SFC overlay and control and orchestration levels. While the management layer is not less important, further research is motivated in deploying management applications for SFC. An example of management applications can include SFC maintenance tools, SFs' real-time status visualization, resources utilization status, troubleshooting tools, and QoS&QoE assessment and visualization.

OAM requirements for SFC are still open research issues [9]. Indeed, there is a need for tools for checking SFC performance and for tracing. Also, the liveliness of SFs needs to be verified to avoid service failure. Moreover, the management of heterogeneous types of SFs is challenging (different physical and virtual appliances). Furthermore, adapting between SDN-enabled and non-SDN enabled devices, the virtualized and non-virtualized infrastructures is challenging.

In order to deploy SFC, some techniques require SF modifications, support for certain protocols, which is not always applicable, and add complexity to SFC deployment. Thus flexible SFC solutions, taking into consideration the current state of SFs, reducing requirements and support of protocols, will simplify the SFC deployment process and integration to production networks and SFC management.

While a network can be managed by different operators (if there is a trust relationship), policy conflicts may raise other concerns since SFCs are a representation of a policy, and different operators may not agree on the same policy. Furthermore, Over The Top (OTT) Service Providers claim another level of service chaining management that is not taken into consideration by ISPs. Thus, another level of SFC management is requested to emphasize user prioritization (e.g., free or premium users), the related resources management and security issues. The authors in [10] highlight some of the open issues in SFC deployment for OTT service providers.

### QUALITY OF SERVICE

One of the challenging issues of SFC is the QoS for deploying SFCs and delivering complex services. With the increasing constraints of hyper-connectivity in next-generation networks, the critical industry requirements and incremental users demands, the QoS requirements grow as a result. While QoS for SFC incurs different SFC problems, some challenges are directly related to QoS. QoS can be assessed by several metrics to identify the performance of traffic flows and services. Metrics such as bandwidth, throughput, delay, packet loss ratio, latency and service availability directly impact the SFC QoS. Yet, the combination of different metrics is a complex problem and different objectives can be contradictory.

Along with QoS, the Quality of Experience (QoE) needs to be considered as well. QoE reflects the quality of experience in the form of satisfying functional requirements of end-users. Thus, it is important to respect QoS while maximizing QoE which is a challenging problem. As such, QoS and QoE requirements can be contradictory. Indeed, respecting QoS does not necessarily require respecting QoE, for example, QoS requirements for low resource consumption can reduce QoE by possible service failure and unavailability. Also, optimizing latency and computational resources does not necessarily improve QoE. Furthermore, there is a need to assess QoS and QoE satisfaction in highly dynamic environments and retrieve a tradeoff between them. Dynamically analyzing and evaluating QoS and QoE, and adapting them to network changes, is an interesting challenge to cope with.

### SECURITY

Security is the worry of every network infrastructure provider, starting with providers of enterprise networks, datacenters and up to carrier networks. Basically, a secure SFC process inherits various

> In the case of a network managed by different operators, security remains the main concern. In the case of a direct peering between the networks, the encapsulation and tunnels have to be used. Assuming the operators trust each other, the SFC information should be further encrypted in order to prevent third-parties from modifying packets and bypassing the policy.

functionalities from the operator's security policies based on risk analysis. However, secure SFC requires operations at different components of the SFC-enabled domain. Indeed, the boundaries of an SFC domain must implement filters for the traffic steering protocols and transport encapsulation protocols used along with continuous audits. The classification process is also critical for the security of SFC requiring the implementation of the authentication and checkups beforehand. Also, the SFC components set as trusted devices are critical for the SFC operation's security. If such devices are compromised, the full chains are compromised. Hence, encryption should be used with transport protocols to ensure information integrity and confidentiality. Furthermore, if the operator's security policies are not taken into account, several attacks may occur such as spoofing, DDoS, reflection, insertion and SFP manipulation.

While some SFC techniques can raise security concerns and threaten infrastructure security, the infrastructure security policies can block some SFC operations. In order to achieve SFC overlay, the SFC information is shared between the SFC elements, usually based on an SFC header, packet fields or tags. Such information can be forged and the traffic can be manipulated. Even though SFC is expected to be applied in a single administrative domain, the risk of confidentiality and integrity is high. Moreover, most of the SFC proposals do not include encryption mechanisms to secure the SFC information and packets information (besides [11] that proposes authenticated and encrypted NSH service chains).

While some traffic steering schemes often assume transparent SFs [1], some opaque SFs in the chain can modify the packet headers. Being unaware of such modifications can lead to inconsistency in the SFP or convey misleading policies. In other words, when SFs do not preserve the SFC information, the modifications of the packet header (mainly the modification of the five tuples or SFC header) results in losing the SFC information; consequently, the SFC process is broken. Moreover, both opaque and transparent SFs should be considered in the design of SFC solutions.

In the current state of SFC, some SFC headers are not supported by security middleboxes, mainly inter-domain SFC. Therefore, the traffic with an SFC header can be interpreted as suspicious or unrecognized. Thus, SFC traffic can be blocked or dropped. Even the SFC solutions based on existing packet fields can be considered as suspicious traffic, for example, the modified MAC addresses (encoding some SFC information) can be detected as non-legitimate traffic.

In the case of a network managed by different operators, security remains the main concern. In the case of a direct peering between the networks, the encapsulation and tunnels have to be used. Assuming the operators trust each other, the SFC information should be further encrypted in order to prevent third-parties from modifying packets and bypassing the policy. Also, the tunnels should be protected properly.

### SFC IMPLEMENTATIONS AND RESEARCH DIRECTIONS

Although intensive works have been achieved in SFC, there are still several gaps and open issues. Besides the support of SFC in some open-source SDN controllers (OpenDaylight, ONOS), NFV orchestrators (OpenStack) and some open source switches (OpenVswitch, FD.io VPP), there is still a need for programming more SFC functionalities (e.g., partial symmetry, SFs discovery, SFC controller discovery, testing, troubleshooting, analytics, security checks and other OAM functionalities).

Although SDN and NFV supported the SFC deployment recently, network slicing and Fog computing can further empower the SFC deployment. Moreover, edge computing is another new technology that can be used for SFC to reduce latency and improve the user experience. Applying cognitive computing at the network edge can provide dynamic and elastic storage and computing services [12]. While SFC is being involved in different networks such as 5G, mobile and Internet of Things (IoT) networks, very few papers are published about these topics [13]–[15].

### CONCLUSION

In this article, we have discussed different research problems in SFC and related challenges. The challenges are grouped into different active research areas in SFC. Some of the challenges are originated from the SFC concept such as the problem of TS and SFC management. On the other hand, other problems are influenced by other technologies that are still under research such as SDN and NFV. These problems impact SFC, as one of their application fields. This includes service allocation and provisioning, SFC composition and path selection. Other problems also affect SFC, mainly related to networking in general, such as the quality of service and security problems. The discussion in this article does not only show that there is much work to be done in the SFC area but also in the related technologies including SDN and NFV.

### ACKNOWLEDGMENT

This work was partially supported by the Grant Project ITIC-TRANSPORT, Moulay Ismail University of Meknes. It was also partially supported by the European Union's Horizon 2020 Research and Innovation Program through the MonB5G Project under Grant No. 871780. This work was also supported in part by the Academy of Finland 6Genesis project under Grant No. 318927 and by the Academy of Finland CSN project under Grant No. 311654.

## BiogRaphiES

Hajar Hantouti (h.hantouti@edu.umi.ac.ma) received her M.S. degree in information systems security from the University of IbnTofail, Morocco, in 2014. She is currently a Ph.D. candidate at the University of Moulay Ismail, Morocco. Her research interests include service function chaining, software-defined networking and network function virtualization.

Nabil Benamar (n.benamar@est.umi.ac.ma) is currently a professor of computer sciences at the School of Technology, Moulay Ismail University, Morocco. He is an IPv6 expert and consultant with many international organizations. His research interests are Software Defined Networks, vehicular networks, DTNs, ITS, IPv6 and IoT.

Tarik Taleb (talebtarik@ieee.org, tarik.taleb@aalto.fi) is currently a professor at the School of Electrical Engineering, Aalto University, Finland, and the Centre for Wireless Communications (CWC), University of Oulu, Finland. He has worked as a senior researcher and 3GPP standards expert at NEC Europe Ltd. Prior to his work at NEC, until March 2009, he worked as an assistant professor at the Graduate School of Information Sciences, Tohoku University, Japan. His research interests lie in the field of architectural enhancements to mobile core networks (particularly 3GPP's), mobile cloud networking, mobile multimedia streaming, and social media networking. He has also been directly engaged in the development and standardization of the Evolved Packet System. He is a member of the IEEE Communications Society Standardization Program Development Board and serves as Steering Committee Chair of the IEEE Conference on Standards for Communications and Networking.